\documentclass[aps,prb,twocolumn,floatfix,superscriptaddress,showpacs,
showkeys,amssymb,amsmath]{revtex4}
\usepackage{graphicx}
\usepackage{amssymb}

\begin{document}
\title{Level structure and spin-orbit effects in semiconductor nanorod dots}
\author{C.\ L.\ Romano}
\affiliation{Department of Physics and Astronomy and Nanoscale and
Quantum Phenomena Institute, Ohio University, Athens, Ohio
45701-2979} \affiliation{Department of Physics, University of
Buenos Aires, Ciudad Universitaria, Pab.\ I, C1428EHA Buenos
Aires, Argentina}
\author{S.\ E.\ Ulloa}
\affiliation{Department of Physics and Astronomy and Nanoscale and
Quantum Phenomena Institute, Ohio University, Athens, Ohio 45701-2979}
\author{P.\ I.\ Tamborenea}
\affiliation{Department of Physics, University of Buenos Aires,
Ciudad Universitaria, Pab.\ I, C1428EHA Buenos Aires, Argentina}
\date{\today }

\begin{abstract}
We investigate theoretically how the spin-orbit Dresselhaus and Rashba
effects influence the electronic structure of quasi-one-dimensional
semiconductor
quantum dots, similar to those that can be formed inside semiconductor nanorods.
We calculate electronic energy levels, eigen-functions, and effective $g$-factors
for coupled, double dots made out of different materials, especially GaAs and InSb.
We show that by choosing the form of the lateral confinement, the contributions
of the Dresselhaus and Rashba terms can be tuned and suppressed, and we consider
several possible cases of interest.
We also study how, by varying the parameters of the double-well confinement
in the longitudinal direction, the effective $g$-factor can be controlled to a large
extent.

\end{abstract}

\maketitle

\section{INTRODUCTION}

In recent years, much of the research in semiconductor physics has
been shifting towards {\em spintronics}, \cite{pri,wol-aws-buh} the novel
branch of electronics in which the information is carried, at
least in part, by the spin of the electrons.
The electron spin might be used in the future to build quantum computing
devices combining logic and storage based on spin-dependent effects in
semiconductors.
In order to achieve this goal, much study has been devoted recently to
magnetic and optical \cite{opt} properties of semiconductors quantum
dots \cite{tav,los-div} and quantum wells. \cite{gol}
One of the most popular spin-based devices was proposed by Datta
and Das.\cite{dat-das}
Improvements to the original design have been proposed recently by
Egues et al.\ \cite{egu-bur-los,sch-egu-los}
The Datta-Das device makes use of the Rashba spin-orbit coupling \cite{ras,byc-ras}
in order to perform controlled rotations of the spins of electrons
passing through the channel of a field-effect transistor (FET),
thus creating a spin-FET.
The Rashba term is the manifestation of the spin-orbit interaction
in quasi-one-dimensional (quasi-1D) semiconductor nanostructures lacking
{\it structural} inversion symmetry.
Additionally, the lack of {\it bulk} inversion symmetry enables another
spin-orbit term in the electronic Hamiltonian, the Dresselhaus term, \cite{dre}
which is also taken into account in the spin-FET design introduced
in Ref.\ [\onlinecite{sch-egu-los}].

The influence of the Rashba and Dresselhaus Hamiltonians in quantum
dots (QD) has recently been treated in a number of theoretical
works.
The most-often studied geometry is that of quasi-two-dimensional
dots with parabolic confinement in the plane.\cite{gov,tsi-loz-gog,des-ull-mar}
On the other hand, there is a growing interest and experimental progress
in another type of quantum dots defined inside quasi-1D
structures called nanorods or nanowhiskers.\cite{lie}
In these structures, additional confinement in the longitudinal direction
can be introduced with great precision, thereby allowing the formation of
quasi-1D heterostructures, such as multiple quantum
dots \cite{sam,bjo-etal} and dot superlattices. \cite{wu-fan-yan}
Nanorods can be grown out of numerous semiconductor materials.
Their lateral widths can be controlled by selecting the size of the
gold nanoparticles which are used to catalyze their growth and
can be made as small as 3 nm.\cite{mor-lie}
Recently, the transport properties of these nanorod dots have been
measured and the gated control of the number of electrons in them has
been demonstrated.\cite{bjo-etal}

Motivated by this experimental progress, we study theoretically the
electronic structure of quasi-1D coupled double dots including spin-orbit
effects.
This type of dot systems has also attracted interest in the
field of quantum control of orbital wave functions due to their simplicity
and tunability.\cite{tam-met,zha-zha,chaos,cre-pla}
As we will see here they are also well-suited for applications
involving control of the spin degrees of freedom since they allow
a great deal of control over the Rashba and Dresselhaus Hamiltonians.
In this paper we study the influence of the Rashba and Dresselhaus
spin-orbit Hamiltonians on the electronic structure of quasi-1D QDs,
akin to those formed in semiconductor nanorods.
Our emphasis on the spin-orbit interaction is obviously motivated by
the current widespread interest in developing spintronic applications,
which require a detailed understanding of the dynamics of the spin degree
of freedom in semiconductor nanostructures.

Let us denote by $x$ and $y$ the two transversal and by $z$ the longitudinal
direction of a quasi-1D nanorod, and let us call $V_z(z)$ the confining
potential that defines a pair of coupled QDs along the nanorod.
The laterally-confining potentials $V_x(x)$ and $V_y(y)$ are crucial
in the determination of the Rashba and Dresselhaus Hamiltonians
and we consider different combinations of these potentials which
can arise in our elongated geometry.
We calculate the energy spectra and the wave functions by exact
numerical diagonalization of the total Hamiltonian and analyze
how the energy levels and the effective $g$-factor change as the Rashba
and Dresselhaus couplings are modulated by varying the lateral
confining potentials.
Furthermore, we study the effect of varying the size of one of the
dots and the width of the central barrier between them.
Since the strength of the spin-orbit interaction varies greatly
among semiconductor compounds, we look at several materials
such as GaAs, InSb, GaSb, and InAs.
Finally, we investigate the effective spin $\left\langle S_{z} \right\rangle$
as a function of the strength of the Rashba-like term for all the
eigenfunctions of InSb with two different geometries.

The quantization along different directions results in peculiar
spin-momentum dependence.
This in turn results in SO effects that depend strongly on the
symmetries of the lateral confinement potentials.
As such, the observation of SO spin splittings, as we will see,
is directly attributable to asymmetry of the confinement and
provides an interesting probe of built-in strain fields and/or
unbalanced composition gradients.

We organize the paper as follows.
In Sec.\ \ref{sec:intro} we introduce the effective one-dimensional Hamiltonian
and list the simplified forms it takes depending on the choice of
confinement potentials.
In Sec.\ \ref{sec:energies} we present the results for the energy levels
including either the Dresselhaus or the Rashba term.
In Sec.\ \ref{sec:gfactor} we study the effective {\em g}-factor and the
expectation value of the {\em z}-component of the spin as a function of
the strength of the Rashba term for different semiconductors and eigenstates.
In Sec.\ \ref{sec:conclusions} we provide a discussion and conclusion.

\section{The one-dimensional Hamiltonian}
\label{sec:intro}

\begin{figure}[tbp]
\includegraphics*[width=8cm]{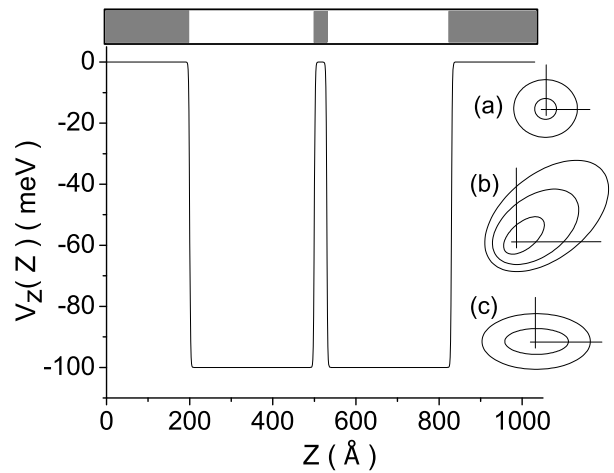}
\caption{Potential-energy profile and schematic drawing of two Al/InSb
coupled nanorod quantum dots.
For InSb-based systems we take a well height of 100 meV,
and for Al/GaAs, 220 meV. In this example, the QD width is $300
\mbox{ \AA}$ and barrier width 30 \AA, with smoothly changing
barriers over a width of a few angstroms.
The drawings (a)-(c) illustrate the lateral confinement
geometries described in the text.}
\label{fig:potential_dots}
\end{figure}

We start with the complete Hamiltonian for a three-dimensional
semiconductor structure in the absence of magnetic field,
\begin{equation}
H = \frac{p^2}{2m^{*}} + V(\mathbf{r}) + H_D + H_R,
\end{equation}
where $m^*$ is the conduction-band effective mass,
$\mathbf{p}$ is the momentum, $V(\mathbf{r})$ is the confinement
potential, and $H_D$ and $H_R$ are the general Dresselhaus
and Rashba Hamiltonians.\cite{des-ull-mar-2004b}
Here we follow the current practice of calling Rashba terms
those spin-orbit contributions to the Hamiltonian that arise
due to the structural inversion asymmetry of the nanostructure,
as opposed to the Dresselhaus terms which come from the bulk
inversion asymmetry of the III-V semiconductors.
Integrating out the {\em x} and {\em y} variables, we obtain the
following effective one-dimensional Hamiltonian:
\begin{equation}
H_{1d} = \frac{p_{z}^{2}}{2m^{*}}+V_{z}\left(z\right)+H_{1dD}+H_{1dR},
\end{equation}
\begin{eqnarray}
H_{1dD}&=&\frac{\gamma_{D}}{\hbar^{3}} \{\sigma_{x}\left\langle
p_{x}\right\rangle \left(\left\langle p_{y}^{2}\right\rangle
-p_{z}^{2}\right) -\sigma_{y}\left\langle p_{y}\right\rangle
\left(\left\langle p_{x}^{2}\right\rangle -p_{z}^{2}\right)\nonumber\\
&+& \sigma_{z}p_{z}\left(\left\langle p_{x}^{2}\right\rangle
-\left\langle p_{y}^{2}\right\rangle \right)\}, \\
H_{1dR} &=& \frac{\gamma_{R}}{\hbar}\{\sigma_{x}\left(\left\langle
\frac{\partial V_y}{\partial y}\right\rangle p_{z}-\frac{\partial
V_z}{\partial z}\left\langle p_{y}\right\rangle
\right)- \nonumber \\
&-& \sigma_{y}\left(\left\langle \frac{\partial V_x}{\partial x}\right\rangle
p_{z}-\frac{\partial V_z}{\partial z}\left\langle
p_{x}\right\rangle \right)+ \nonumber \\
&+& \sigma_{z}\left(\left\langle \frac{\partial V_x}{\partial
x}\right\rangle \left\langle p_{y}\right\rangle-\left\langle
\frac{\partial V_y}{\partial y}\right\rangle \left\langle
p_{x}\right\rangle \right)\},
\end{eqnarray}
where $\sigma_i$, $i=x,y,z$, are the Pauli matrices,
$H_{1dD}$ is the one-dimensional Dresselhaus term,
and $H_{1dR}$ is the Rashba-like term enabled by the
inversion asymmetry of the laterally confining potentials $V_x$ and
$V_y$.
$\gamma_{R}$ and $\gamma_{D}$ are parameters that depend on the
materials.
The averages $\langle \ldots \rangle$ are taken over the lowest-energy
wavefunctions of the laterally confining potentials as we assume small
nanorod widths.
In Table I we present the parameters used in our calculations for
different semiconductor materials.
An example of the confining potential in the longitudinal direction,
$V_{z}(z)$, is shown in Fig.\ \ref{fig:potential_dots}, along with a
schematic drawing of the nanorod QDs.

We now list four different possibilities for the confining potentials
$V_x(x)$ and $V_y(y)$, based on the degree of symmetry of the structure.
The Dresselhaus and Rashba Hamiltonians simplify considerably due to the
fact that, in the absence of a magnetic field, the eigenstates can
be chosen real, and therefore, expectation values of the momentum are
zero.\cite{sha}
The four cases are (see Fig.\ 1(a)-(c) for schematic drawings of
the potentials in the first three cases):

\medskip
\noindent
{\bf (a)} Circular: $V_{x}\left(x\right), V_{y}\left(y\right)$ have
inversion symmetry about the origin and are equal,
$V_{x}\left(x\right)=V\textrm{$_{y}$}\left(y\right)$:
\begin{equation}
H_{SO} = H_{1dD}+H_{1dR}=0,
\end{equation}
yields no SO contributions.

\medskip
\noindent
{\bf (b)} $V_{x}\left(x\right), V_{y}\left(y\right)$ have no inversion symmetry
but are equal, $V_{x}\left(x\right) = V_{y}\left(y\right)$:
\begin{equation}
H_{SO} = H_{1dD}+H_{1dR}=
\frac{\gamma_{R}}{\hbar}\left\langle
\frac{\partial V_x}{\partial x}\right\rangle
p_{z}\left(\sigma_{x}-\sigma_{y}\right),
\end{equation}
so that only Rashba terms are present.

\medskip
\noindent
{\bf (c)} Elliptical: $V_{x}\left(x\right), V_{y}\left(y\right)$ are
inversion symmetric functions and different, $V_x(x) \neq V_y(y)$:
\begin{equation}
H_{SO} = H_{1dD} + H_{1dR}=
\frac{\gamma_{D}}{\hslash^{3}}\sigma_{z}p_{z}\left(\left\langle
p_{x}^{2}\right\rangle -\left\langle p_{y}^{2}\right\rangle\right),
\end{equation}
results in only Dresselhaus terms.

\medskip
\noindent
{\bf (d)} $V_{x}\left(x\right), V_{y}\left(y\right)$ have no
inversion symmetry and are different,
$V_{x}\left(x\right)\neq V_{y}\left(y\right)$:
\begin{widetext}
\begin{eqnarray}
H_{SO}=H_{1dD}+H_{1dR}=\frac{\gamma_{D}}{\hslash^{3}}\sigma_{z}p_{z}
\left(\left\langle p_{x}^{2}\right\rangle -\left\langle p_{y}^{2}
\right\rangle \right)+\frac{\gamma_{R}}{\hbar} p_{z}
\left(\sigma_{x}\left\langle \frac{\partial V}{\partial y}
\right\rangle - \sigma_{y}\left\langle
\frac{\partial V}{\partial x} \right \rangle \right),
\end{eqnarray}
represents the most general case and both Rashba and Dresselhaus
contributions are present.
\end{widetext}

For the calculation of the effective $g$-factor we introduce
a weak magnetic field along the $z$-direction.
The field is chosen small so that the $x-y$ orbital wave functions
are not perturbed significantly.
Thus, we only add a Zeeman term to the Hamiltonian,
$H_{Z}=\frac{\mu_B}{2} g_0 \, B \sigma_{z}$,
where
$\mu_B$ is the Bohr magneton, \textbf{B} is the
magnetic field, and $g_0$ is the electron $g$-factor as per
Table I.
To calculate the energy levels and eigenfunctions, we expand the
total Hamiltonian on a basis set of $300$ wave functions of the
quantum box of size $L$, i.e.\
$\phi_{n,s}(z)=\sqrt{\frac{2}{L}}sin(\frac{n\pi z}{L})\chi\left(s\right)$,
where $\chi(s)$ is the spin function, and diagonalize it numerically
without further approximations.
The size $L$ of the box is such that the whole double-dot structure
is enclosed in the box, including the barriers on the sides of the dots
and as such is irrelevant in the final eigenstates.
We should notice that the geometry of the dots that we study here
includes widths of 2-5 nm, while the most common nanorod widths
in experiments are of the order of tens of nm.
However, as we mentioned above, there are no experimental limitations
to reducing the nanorod width to values we consider here.
Smaller widths allow us to explore the basic physics and control of
electronic wave functions with only one relevant lateral energy sublevel.
Moreover, notice that typical charge depletion induced by the
free surfaces further reduces the effective width of the nanorods,
making them more 1D-like.
A final comment is that the incorporation of additional transverse
levels in the nanorod is straightforward, but results in systems
of coupled differential equations.

\begin{table}[h]
\caption{Parameters for semiconductors\cite{des-ull-mar-2004b}}
\label{Tab1}\setlength{\belowcaptionskip}{10pt} \centering
\begin{tabular}{ccccc}
\hline Parameter&GaAs&GaSb&InAs&InSb
\\ \hline
$m^{*}=m/m_{0}\cite{tab}$&$0.067$&$0.041$&$0.0239$&$0.013$\\
$\gamma_{R}\cite{tabl}$($A^{2}$)&$5.33$&$33$&$110$&$500$\\
$\gamma_{D}\cite{tab}$($meV/A^{3}$)&$24$&$187$&$130$&$220$\\
$g_{0}$&$-0.44$&$-7.8$&$-15$&$-51$\\ \hline
\end{tabular}
\end{table}

\section{ENERGY LEVELS}
\label{sec:energies}

We present results for the energy levels in cases (b) and (c),
i.e.\ with only Rashba and Dresselhaus terms present,
respectively.
The general case (d) does not present qualitatively different
features from (b) or (c) and therefore we concentrate here on the
simpler cases.
For case (b) we fix the strength of the Rashba term by giving the
structural electric field
$\left\langle \frac{\partial V}{\partial x}\right\rangle$.
For case (c), we use as confining potentials in the lateral directions
two harmonic-oscillator potentials with different frequencies:
$V_{q}(q) = \frac{1}{2} m^{*} \omega_{q}^{2} q^2$,
$q = x, \: y$.
These potentials have associated characteristic lengths
$\ell_{q}=\sqrt{\hbar/m^{*}\omega_{q}}$.

\begin{figure}[tbp]
\includegraphics*[width=7cm]{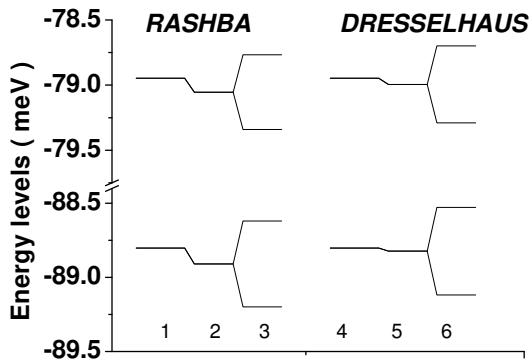}
\caption{
Ground-state and first-excited-state energy levels of the InSb nanorod
QDs shown in Fig.\ 1.
We compare the eigenenergies of
(1,4) $H_{0}= \frac{P_{z}^{2}}{2m^{*}} + V_{z}(z)$ to those of
(2) $H_{0}+H_{1dR}$,
(5) $H_{0}+H_{1dD}$,
(3) $H_{0}+H_{1dR}+H_{Z}$, and
(6) $H_{0}+H_{1dD}+H_{Z}$.
$B=0.2 \mbox{T}$.  }
\label{fig2}
\end{figure}

In Fig.\ \ref{fig2} we plot the two lowest energy levels for the
InSb QDs taking $\left\langle \frac{\partial V}{\partial
x}\right\rangle= 0.5 \, \mbox{meV}/\mbox{ \AA}$ for case (b), and
$\ell_x=50 \, \mbox{\AA}$, $\ell_y=20 \mbox{ \AA}$ for case (c).
The indices on the horizontal axis denote the inclusion of
different terms in the Hamiltonian. The figure shows how the
energy levels of $H_{0}$ (indices 1 and 4) are changed by the
inclusion of a Rashba contribution $H_{1dR}$ (case (b), index 2),
and of a Dresselhaus contribution $H_{1dD}$ (case (c), index 5),
without magnetic field. With a weak magnetic field we have total
Hamiltonians $H_{0}+H_{1dR}+H_{Z}$ (index 3) and
$H_{0}+H_{1dD}+H_{Z}$ (index 6). We have carried out analogous
calculations for the semiconductors quoted in Table I and the
results were qualitatively similar to the ones shown here. The
main general conclusion is that the effect of $H_{1dR}$ is always
stronger than that of $H_{1dD}$ for the chosen parameters, which
are representative of possible experimental situations.
We note that the Rashba and Dresselhaus terms do not remove the spin
degeneracy (as expected from the Kramers degeneracy in the absence
of magnetic field) but that they simply shift the levels downwards,
the strength of the shifts being controlled by the parameters
$\left\langle \frac{\partial V}{\partial x}\right\rangle $ for
Rashba and $\ell_{x}$ and $\ell_{y}$ for Dresselhaus.
For the parameters chosen here the Rashba shift is of the order of $0.1
\mbox{meV}$ for InSb and $0.1 \mu\mbox{V}$ for GaAs while the
Dresselhaus shift is of the order of $0.01 \mbox{meV}$ for InSb
and $0.01 \mu\mbox{V}$ for GaAs.

\begin{figure}[tbp]
\includegraphics*[width=8cm]{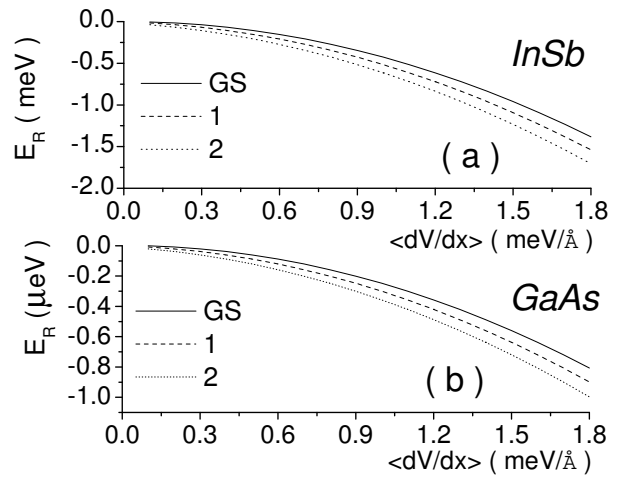}
\caption{
Contribution of the Rashba term to the energy levels of InSb (a) and GaAs (b)
QDs as a function of $\left\langle \frac{\partial V}{\partial x}\right\rangle$.
GS: Ground state, $1$ and $2$: first and second excited states, respectively.
Notice effect is much smaller in GaAs (energy given in $\mu\mbox{eV}$),
as anticipated.}
\label{fig3}
\end{figure}

\begin{figure}[tbp]
\includegraphics*[width=7cm]{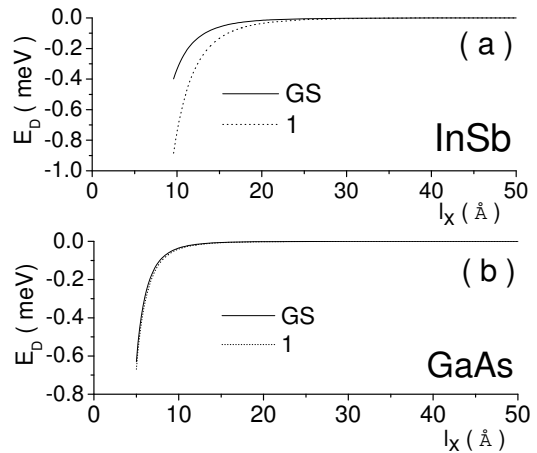}
\caption{Contribution of the Dresselhaus term to the energy levels
of InSb as a function of $\ell_{x}$ for the ground state (GS) and
the first excited state (1) for $\ell_{y}=50 \mbox{ \AA}$.
Level splitting in GaAs is barely visible on the same scale as in
InSb.}
\label{fig4}
\end{figure}

As can be seen in Fig.\ \ref{fig3} the energy shift produced by
the $H_{1dR}$ varies quadratically with the structural electric
field $\left\langle \frac{\partial V}{\partial x}\right\rangle$.
In Fig.\ \ref{fig4} we show how the energy levels vary in case (c)
as a function of $\ell_{x}$ for the two lowest-energy states for
fixed $\ell_{y}=50\mbox{ \AA}$. The functional dependence here is
also parabolic. This suggests that the spin-orbit corrections to
the energy levels could be calculated fairly accurately with
second-order perturbation theory. We performed the second-order
perturbative calculation in the case with Rashba Hamiltonian, with
a small magnetic field applied (0.1 T) in order to work with
non-degenerate perturbation theory. A comparison between the exact
and second-order energies shows, for example, a difference of 17\%
for $\left\langle \frac{\partial V}{\partial x}\right\rangle = 1.5
\, \mbox{meV}/\mbox{\AA}$, and increasing differences for larger
Rashba fields, as expected. These results agree qualitatively with
those of Ref.\ [\onlinecite{tsi-loz-gog}] for quasi-2D circular dots,
where differences of up to 30\% between the results of exact
calculations and of second-order perturbation theory have been
found.

\section{Effective $g$-factor}
\label{sec:gfactor}

\begin{figure}[tbp]
\includegraphics*[width=8cm]{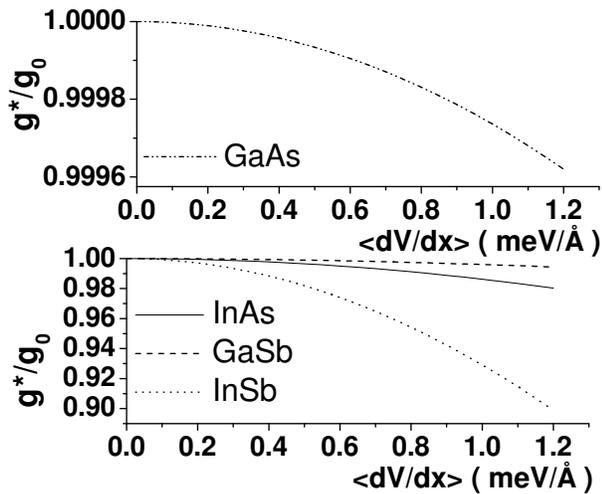}
\caption{
Effect of the Rashba Hamiltonian on the effective $g$-factor.
$g^*/g_{0}$ for the ground state for different semiconductors as a function of
$\left\langle \frac{\partial V}{\partial x}\right\rangle$.}
\label{fig5}
\end{figure}

The small magnetic field
${\mathbf B} = 0.1 \mbox{T} \, {\mathbf z}$
breaks the spin degeneracy of the ground state and allows the calculation
of the effective $g$-factor ($g^*$) as a function of
$\left\langle \frac{\partial V}{\partial x}\right\rangle$ (case (b))
for GaAs, InSb, InAs and GaSb.
In the figures we report normalized $g$-factors:
\begin{equation}
\frac{g^*}{g_0}=\frac{\left(E_{2}-E_{1}\right)}{\frac{\mu_B B g_0}{2}},
\end{equation}
where $E_1$ and $E_2$ are the Zeeman-split ground-state levels.
Figure \ref{fig5} shows the results for case (b) (i.e. with only Rashba
contributions) as a function of
$\left\langle \frac{\partial V}{\partial x}\right\rangle$.
The decreasing trend of $g^*$ is qualitatively similar for all the materials
but the magnitude of this Rashba effect varies greatly among them.
The decrease of the $g^*$ is strongest for InSb and weakest for GaAs.

\begin{figure}[tbp]
\includegraphics*[width=8cm]{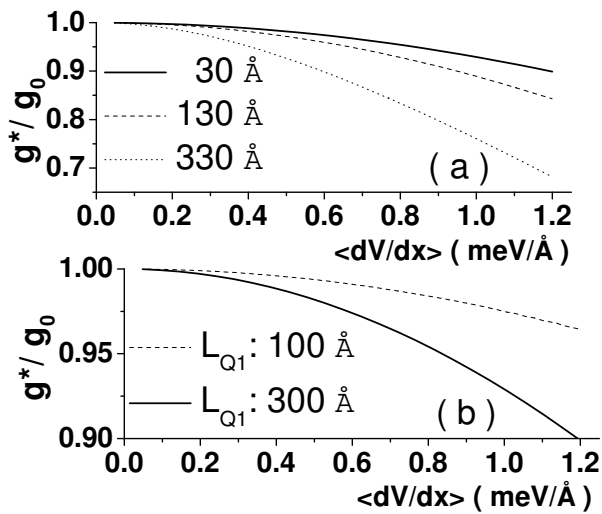}
\caption{Normalized effective $g$-factor for the ground state of
InSb structures with Rashba Hamiltonian. (a) For different barrier
widths $w=30, 130, 330 \mbox{\AA}$. (b) For different sizes of the
QDs. Asymmetric case: $L_{QD1}=100 \mbox{ \AA}$ and
$L_{QD2}=500\mbox{ \AA}$; symmetric case $L_{QD1}=L_{QD2}=300
\mbox{ \AA}$.} \label{fig6}
\end{figure}

We now examine what happens to $g^*$ when one modifies the
features of the longitudinal potential $V_z(z)$, such as
 the barrier width $w$ and the size of the QDs (so far we
have taken $L_{QD1}=L_{QD2} =300 \mbox{ \AA)}$. In Fig.\
\ref{fig6}(a) we show $g^*$ for $w=30, 130$, and $330 \mbox{ \AA}$
as a function of $\left\langle \frac{\partial V}{\partial
x}\right\rangle$. We increase the barrier width but reducing at
the same time the sizes of the two QDs so that the total size of
the structure remains constant at $630 \mbox{ \AA}$. We note that
increasing $w$ leads gradually to having two uncoupled QDs and to
a stronger variation of $g^*$. In Fig.\ \ref{fig6}(b) we set $w=
30 \mbox{ \AA}$ and change the QDs' sizes. We take $L_{QD1}= 100
\mbox{ \AA}$ and $L_{QD2}= 500\mbox{ \AA}$ in one case, and
$L_{QD1}=L_{QD2}= 300\mbox{ \AA}$ in the other. We observe here
that the {\em symmetric} potential produces a stronger variation
of $g^*$ than the {\em asymmetric} one.

We look at these symmetric and asymmetric structures in more
detail, and calculate the expectation value $\left\langle
S_{z}\right\rangle$ as a function of $\left\langle \frac{\partial
V}{\partial x}\right\rangle$ for InSb dots and for the four lowest
pairs of states (Zeeman doublets). Again a magnetic field
$\mathbf{B} =0.1 \mbox{T}$ is included. As expected, $\left\langle
S_{z}\right\rangle = \pm\frac{1}{2}$ in the absence of
$\left\langle \frac{\partial V}{\partial x}\right\rangle$. Figure
7 shows the results for a symmetric structure with
$L_{QD1}=L_{QD2}=300 \mbox{ \AA}$ and Fig.\ 8 for an asymmetric
one with $L_{QD1}=100 \mbox{ \AA}$ and $L_{QD2}=500 \mbox{ \AA}$.
The symmetric case shows a crossing in $\left\langle S_{z}\right\rangle$
(Fig.\ 7(a)) while the asymmetric one does not (Fig. 8(a)).
Using this information we recalculate the effective $g$-factor for the first
four pairs of eigenstates for the symmetric (Fig.\ 7(b)) and asymmetric
(Fig.\ 8(b)) structures.
The effective $g$-factor, given here by the difference in
$\left\langle S_{z}\right\rangle$ values for every Zeeman pair,
vanishes at the crossing of $\left\langle S_{z}\right\rangle$.
This vanishing of $g^*$ is a potentially useful effect in spintronics
applications, as it can be achieved as a function of the potentially
adjustable Rashba parameter
$\left\langle \frac{\partial V}{\partial x}\right\rangle$.
It is interesting to note how different spatial asymmetry, introduced by
the confinement potential along $z$ (i.e.\ different size dots),
has strong effect on $g^*$, and results in a finite value even
at large Rashba fields.

\begin{figure}[tbp]
\includegraphics*[width=8cm]{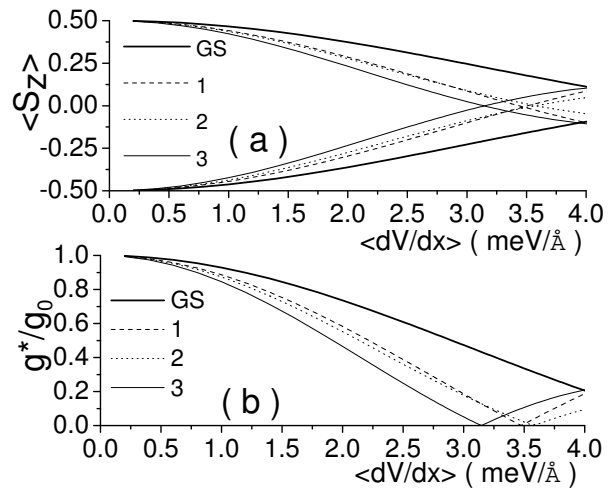}
\caption{Mean value of $S_z$ and effective $g$-factor for InSb
systems with symmetric $V_z(z)$ (two equal dots with
$L_{QD1}=L_{QD2}=300 \mbox{ \AA}$). (a) $\left\langle
S_{z}\right\rangle$ as a function of
    $\left\langle \frac{\partial V}{\partial x}\right\rangle$ for the four
lowest-energy doublets (pairs of Zeeman-split states).
(b) $g^*/g_{0}$ for the same states.}
\label{fig7}
\end{figure}

\begin{figure}[tbp]
\includegraphics*[width=8cm]{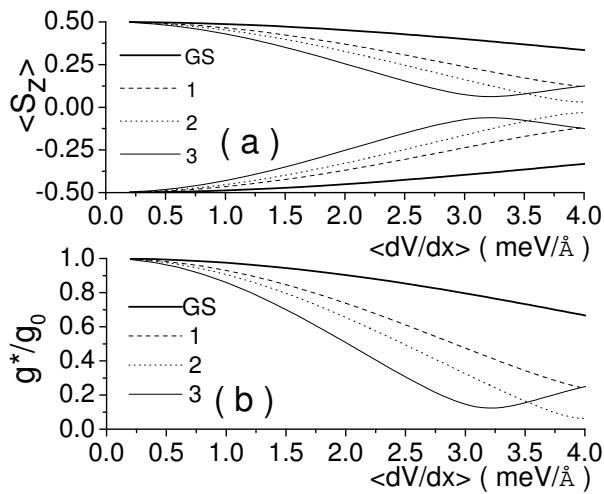}
\caption{Same as Fig.\ 7  for asymmetric $V_z(z)$ with
$L_{QD1}=100 \mbox{ \AA}$ and $L_{QD2}=500 \mbox{ \AA}$}
\label{fig8}
\end{figure}

\section{Conclusions}
\label{sec:conclusions}

We have studied how the spin-orbit Rashba and Dresselhaus
terms modify the electronic structure of nanorod quasi-one-dimensional
double quantum dots.
We have solved the problem by numerical diagonalization of the total
Hamiltonian for varying confining potentials, in the lateral as well as
in the longitudinal directions.
The main conclusions of our work are the following: \\
(1) For our system, the Rashba and Dresselhaus Hamiltonians shift downwards
the energy levels but do not break the spin degeneracy of the electronic levels
in the absence of an external magnetic field (as prescribed by the Kramers
degeneracy.) \\
%This contrasts with the degeneracy-breaking effect that the Rashba term has
%in two- and one-dimensional systems.
%
(2) The Rashba effects are in general stronger than the Dresselhaus effects,
but the latter are not negligible in general either.\\
(3) Changing the strength of the spin-orbit terms, which is done by changing
the lateral confinement length
$\ell_{x}$ or $\ell_{y}$
in the case of Dresselhaus
or the structural electric field
$\left\langle \frac{\partial V}{\partial x}\right\rangle$
in the case of Rashba,
results in energy levels that vary nearly quadratically with the control parameter.
This indicates that the SO corrections to the energy levels are close to
the second-order corrections in perturbation theory.
We verified this result by comparing the exact and the perturbatively
calculated energies.\\
(4) By changing the strength of the Rashba term, the size of the central barrier,
and the size and symmetry of the two QDs, it is possible to control the value of the effective
$g$-factor, which determines the Zeeman splitting.
In particular, it is possible to make the effective $g$-factor equal to zero.

\begin{acknowledgments}
We acknowledge support from the CMSS Program at Ohio University,
Proyectos UBACyT 2001-2003 and 2004-2007, Fundaci\'on Antorchas,
ANPCyT grant 03-11609, and NSF-CONICET through a US-Argentina-Brazil
collaboration grant NSF 0336431.
P.I.T.\ is a researcher of CONICET.
\end{acknowledgments}

\end{document}